\documentclass[a4paper]{jpconf}

\usepackage[normalem]{ulem}
\usepackage{dsfont}
\usepackage[T1]{fontenc}
\usepackage{amsmath}
\usepackage{amssymb}
\usepackage{amsthm}
\usepackage[ansinew]{inputenc}
\usepackage{enumerate}
\usepackage{fancyhdr}
\usepackage{cite}
\usepackage{color}
\usepackage{ifpdf}
\usepackage{graphicx}
\usepackage{rotating}
\newtheorem{theorem}{Theorem}

\newcommand{\R}        {\mathds{R}}

\begin{document}

\title{Late-time behaviour of the Einstein-Vlasov system with Bianchi I symmetry}

\author{Ernesto Nungesser \\ 
Max-Planck-Institut f\"ur Gravitationsphysik\\
Albert-Einstein-Institut\\
Am M\"uhlenberg 1\\
14476 Potsdam, Germany}
\ead{ernesto.nungesser@aei.mpg.de}

\begin{abstract}

The late-time behaviour of the Einstein-dust system is well
understood for homogeneous spacetimes. For the case of Bianchi I
we have been able to show that the late-time behaviour of the Einstein-Vlasov system
is well approximated by the Einstein-dust system assuming that one is close to the unique stationary solution which is the attractor of the Einstein-dust system.
\end{abstract}

\section{Introduction}

The Einstein-Vlasov system with Bianchi symmetry is a system of
integro-differential equations on a Lorentzian manifold $M$ which has
the structure $M=I \times G$ where $I$ is an interval and $G$ a Lie
group. It can be seen as a kind of transport equation, say $Xf=0$, coupled to other
ordinary differential equations. There exists certain limiting process
such that the transport equation is transformed into an ordinary
differential equation. This happens if the function $f$ becomes a
distribution. The resulting system is called Einstein-dust system and
the late-time behavior is well understood. It has been studied by the
theory of dynamical systems and for almost all Lie groups the
$\omega$-limit set is known \cite{WE}. The strategy to study the
late-time behaviour of the Einstein-Vlasov system consists in trying
to show that the late-time behavior is well approximated by the
Einstein-dust system. Physically this makes sense since one might
expect a decay of the velocity dispersion for expanding
cosmological models. The result presented in the following shows that
this is true for case of the Lie group ${\R}^3$ and assuming that one
is close to the unique stationary solution which is the attractor of
the Einstein-dust system.\\

\section{Einstein-Vlasov system with Bianchi I symmetry}
\subsection{Einstein equations with Bianchi I symmetry}
A Bianchi spacetime is defined to be a spatially homogeneous spacetime whose isometry group possesses a three-dimensional subgroup $G$ that acts simply transitively on the spacelike orbits. They can be classified by the structure constants of the Lie algebra associated to the Lie group. We will only consider the simplest case where the structure constants vanish, i.e. the case of Bianchi I where the metric has the following form using Gauss coordinates:
\begin{eqnarray}\label{me}
 ^4 g=-dt^2+g_{ab}(t)dx^adx^b.
\end{eqnarray}
In terms of coordinate expressions
\begin{eqnarray*}
   \rho&=&T^{00}\\
     j_a&=&T_a^0\\
  S_{ab}&=&T_{ab}
\end{eqnarray*}
where $\rho$, $j_a$ and $T_{\mu\nu}$ are the energy density, matter
current and energy momentum tensor respectively.
Using the 3+1 decomposition of the Einstein equations as made in \cite{RA} the basic equations are
\begin{eqnarray}\label{a}
 &&\dot{g}_{ab}=-2 k_{ab}\\
 &&\dot{k}_{ab}=H k_{ab}-2k_{ac}k^c_b-8\pi(S_{ab}-\frac{1}{2}g_{ab}\tr S)-4\pi\rho g_{ab}\\
 &&-k_{ab}k^{ab}+H^2=16\pi\rho\\
 &&T_{0a}=0.
\end{eqnarray}
where $k_{ab}$ is the second fundamental form, $H$ its trace and a dot
above a letter denotes a derivative with respect to the cosmological
(Gaussian) time $t$. It has been assumed that the cosmological constant vanishes.
\subsection{Vlasov equation with Bianchi I symmetry}
For the matter model we will take the point of view of kinetic
theory. This means that we have a collection of particles (in a
cosmological context the particles are galaxies or clusters of
galaxies) which are described statistically by a non-negative
distribution function $f(x^\alpha,p^\alpha)$ which is the density of
particles at a given spacetime point with given four-momentum. We will
assume that all the particles have equal mass (one can relax this condition if necessary, see \cite{PP}) . We want that our matter model is compatible with our symmetry assumption, so we will also assume that $f$ does not depend on $x^a$. In addition to that we will assume that there are no collisions between the particles. In this case the distribution function satisfies the Vlasov equation (See (3.38) of \cite{RA}):
\begin{eqnarray}
\frac{\partial f}{\partial t}+2 k^a_b p^b\frac{\partial f}{\partial p^a}=0.
\end{eqnarray}
where f is defined on the set determined by the equation 
\begin{eqnarray*}
 -(p^0)^2+g_{ab}p^ap^b=-m^2
\end{eqnarray*}
called the mass shell.
For a given Bianchi I geometry the Vlasov equation can be solved explicitly with the result that if $f$ is expressed in terms of the covariant components $p_i$ then it is independent of time. The non-trivial components of the energy momentum tensor are:
\begin{eqnarray}
&& \rho=\int f_{0}(p_{i})(m^2+g^{cd}p_{c}p_d)^{\frac{1}{2}}(\det g)^{-\frac{1}{2}}dp_{1}dp_{2}dp_{3}\\
&& S_{ab}=\int f_{0}(p_{i})p_a p_b(m^2+g^{cd}p_{c}p_d)^{-\frac{1}{2}}(\det g)^{-\frac{1}{2}}dp_{1}dp_{2}dp_{3}\label{cu}
\end{eqnarray}
For this kind of matter all the energy conditions hold. In particular $\rho \geq \tr S \geq 0$.
Our system of equations consists of the equations (\ref{a})-(\ref{cu}).

\section{Central results}

The assumption that the spacetime is close to isotropic is expressed by assuming that the quantity 
\begin{eqnarray}\label{AN}
 F=\frac{\sigma_{ab}\sigma^{ab}}{H^2}.
\end{eqnarray}
is small. Here $\sigma_{ab}$ denotes the trace-free part of the second fundamental form. The quantity $F$ is related to the so called \textit{shear parameter}, which is bounded by the cosmic microwave background radiation and is a dimensionless measure of the anisotropy of the Universe (See chapter 5.2.2 of \cite{WE}).
The other assumption which is needed is that the spacetime is close to ``dust-like''. This is expressed by assuming that the absolute value of the momenta of the particles is bounded. We define $P$ as the supremum of the absolute value of the momenta at a given time $t$:
\begin{eqnarray}\label{P}
 P(t)=\sup \{ \vert p \vert =(g^{ab}p_a p_b)^\frac{1}{2} \vert f(t,p)\neq 0\}
\end{eqnarray}
The main result is the following:

\begin{theorem}
Consider any $C^{\infty}$ solution of the Einstein-Vlasov system with Bianchi I-symmetry and with $C^{\infty}$ initial data. Assume that $F(t_0)$ and $P(t_0)$ are sufficiently small. Then at late times one can make the following estimates:
\begin{eqnarray}
 H(t)&=&-2t^{-1}(1+O(t^{-1}))\\
P(t)&=&O(t^{-\frac{2}{3}+\epsilon})\label{Pp}\\
 F(t)&=&O(t^{-2})
\end{eqnarray}
\end{theorem}
\noindent
where $\epsilon$ is a small and strictly positive constant. These
estimates imply that the spacetime isotropizes and that asymptotically
there is a dust-like behaviour. The proof is based on a bootstrap
argument. See \cite{EN} for the details. Using these estimates one can obtain more detailed information about the behaviour of the metric. In analogy to the Kasner solution one can define the \textit{generalized Kasner exponents} for the non-vacuum case.
Let $\lambda_i$ be the eigenvalues of $k_{ij}$ with respect to $g_{ij}$ we define
\begin{eqnarray}\label{gke}
p_i=\frac{\lambda_i}{H} 
\end{eqnarray}
as the generalized Kasner exponents. Having analyzed carefully the dust case with small data we could conclude that:
\begin{theorem}\label{t2}
 Consider the same assumptions as in the previous theorem. Then
\begin{eqnarray}\label{ke}
 p_i= \frac{1}{3} +O(t^{-1})
\end{eqnarray}
and 
\begin{eqnarray}\label{lp}
g_{ab}=t^{+\frac{4}{3}}[\mathcal{G}_{ab}+O(t^{-2})]\\\label{ll}
g^{ab}=t^{-\frac{4}{3}}[\mathcal{G}^{ab}+O(t^{-2})] 
\end{eqnarray}
where $\mathcal{G}_{ab}$ and $\mathcal{G}^{ab}$ are independent of $t$.
\end{theorem}

\section{Conclusions and Outlook}
The result can be seen as a generalization of theorem 5.4 of
\cite{IS}. We obtain the same the result, but a) we also obtain how
fast the expressions converge b) we obtain an asymptotic expression
for the spatial metric c) we do not assume additional
symmetries. However we used a different kind of restriction namely the
small data assumptions. For the study of more complicated Lie groups a
starting point can be \cite{RT} - \cite{CH} where the late-time asymptotics are obtained assuming an
additional symmetry. Another direction of generalization is the study of the
asymptotics towards the initial singularity. In \cite{SU} and
\cite{SA} the case of two fluids has been studied which leads to the
non-diagonal case. For the Vlasov case already the non-LRS case has
been analyzed in \cite{HU1}. Already in this case surprising new
features like the existence of heteroclinic networks arose. Finally it
would be interesting concerning the direction of the initial
singularity to show that also in the Vlasov case the off-diagonal
components of the metric tend to constants and thus are not important
for the dynamics. See \cite{HN} for the importance of this behaviour and \cite{MK} for consequences in a quantum version.\\

\ack

The author would like to thank Alan D. Rendall and Michael Koehn for a lot of helpful discussions.

\end{document}